\documentclass[prd,preprint,tightenlines,floatfix,showpacs,preprintnumbers,nofootinbib,eqsecnum]{revtex4}

 \usepackage[dvips,final]{graphicx}
  \usepackage{amssymb}
   \usepackage{amsmath}
    \usepackage{amsfonts}
     \usepackage{epsfig}
      \usepackage{bm} 

\begin{document}


\title{Unruh effect for fermions from the Zubarev density operator}

\author{George Y. Prokhorov$^1$}
\email{prokhorov@theor.jinr.ru}

\author{Oleg V. Teryaev$^{1,2,3}$}
\email{teryaev@jinr.ru}

\author{Valentin I. Zakharov$^{2,4}$}
\email{vzakharov@itep.ru}
\affiliation{
{$^1$\sl 
Joint Institute for Nuclear Research, 141980 Dubna,  Russia \\
$^2$ \sl
Institute of Theoretical and Experimental Physics,
B. Cheremushkinskaya 25, 117218 Moscow, Russia \\
$^3$ \sl
National Research Nuclear University MEPhI (Moscow
Engineering Physics Institute), Kashirskoe Shosse 31, 115409 Moscow,
Russia \\
$^4$ \sl
School of Biomedicine, Far Eastern Federal University, 690950 Vladivostok, Russia
}\vspace{1.5cm}
}

\begin{abstract}
\vspace{0.5cm}
Using the Zubarev quantum-statistical density operator, we calculated the corrections to the energy-momentum tensor of a massless fermion gas associated with acceleration. It is shown that when fourth-order corrections are taken into account, the energy-momentum tensor in the laboratory frame is equal to zero at a proper temperature measured by a comoving observer equal to Unruh temperature.
Consequently, the Minkowski vacuum is visible to the accelerated observer as a medium filled with a heat bath of particles with the Unruh temperature, which is the essence of the Unruh effect.
\end{abstract}


\maketitle

\section{INTRODUCTION}

According to the standard formulation of the Unruh effect, an accelerated observer sees Minkowski vacuum state as a thermal bath of particles with a temperature $T_U=\frac{a}{2\pi}$, depending on the proper acceleration $a$ \cite{Unruh:1976db, Davies:1974th,  Fulling:1972md, Crispino:2007eb}. 

The Unruh effect is most easily derived for scalar particles from consideration of the change in the ratio between positive and negative frequency modes of scalar fields in the proper time of the accelerated observer \cite{Unruh:1976db, Crispino:2007eb}. However, the Unruh effect was also established in the general case of theories with arbitrary spin and with interaction on the basis of the algebraic approach \cite{Bisognano:1975ih, Bisognano:1976za}. Also, the Unruh effect was established for fermions in the framework of quantum field theory in terms of path integrals \cite{Unruh:1983ac}.

There were also indications that the Unruh effect may be significant when considering the collisions of elementary particles. In \cite{Castorina:2007eb, Becattini:2008tx} it was shown that the hadronization process can be accompanied by accelerated particle motion, which can be a source of thermalization in elementary processes such as $e^+e^-$ annihilation or $pp$ and $p\bar{p}$ collisions, just due to the Unruh effect, and also allows to explain some features in the multiplicities of hadrons. So in the elementary processes of electron-positron annihilation or proton-(anti)proton collisions, tremendous accelerations can occur, which may  open the way for observing the effects associated with acceleration, while in heavy ion collision also large vorticity may arise, and polarisation in heavy ion collisions provides information both about vorticity and acceleration \cite{Baznat:2017jfj, Rogachevsky:2010ys}.

An interesting new look at the Unruh effect was recently obtained from the standpoint of quantum relativistic statistical mechanics \cite{Becattini:2017ljh, Becattini:2019poj}. In \cite{Becattini:2017ljh} it was shown by calculating the values of quantum correlators for scalar fields at a finite temperature that the average value of any local operator turns out to be zero after subtracting of the vacuum contribution at the proper temperature, measured by a comoving observer, equal to the Unruh temperature. This fact means that the Minkowski vacuum is perceived by the accelerated observer as a medium filled with a thermal bath of particles with a Unruh temperature $\frac{a}{2\pi}$, which is the essence of the Unruh effect.

The analysis in \cite{Becattini:2017ljh} is given for scalar particles. Our main result is a generalization of \cite{Becattini:2017ljh} for the case of massless fermions: we show that for gas of massless fermions with chemical potentials equal to zero, the energy-momentum tensor in the laboratory frame is zero at the proper temperature, measured by a comoving observer, equal to the Unruh temperature.

We use a fundamental approach based on the quantum-statistical density operator of Zubarev \cite{Zubarev, Weert, Buzzegoli:2017cqy, Becattini:2015nva, Buzzegoli:2018wpy} (see also in \cite{Becattini:2019dxo} the current review of Zubarev's approach and its connection with the Kubo formulas, and in \cite{Becattini:2019poj,Becattini:2017ljh} discussion of thermodynamic equilibrium with acceleration, as well as the derivation of the entropy current). The effects associated with acceleration in this operator are described by a term with the boost operator and can be investigated in the framework of quantum field perturbation theory. So, the first and second order corrections in acceleration (and other velocity derivatives) were calculated in \cite{Buzzegoli:2017cqy, Becattini:2015nva, Buzzegoli:2018wpy}, while the third order corrections in axial current were analyzed in \cite{Prokhorov:2018bql}. In particular, it was shown that the standard formula for the Chiral Vortical Effect is reproduced  \cite{Buzzegoli:2017cqy, Buzzegoli:2018wpy} and corrections to it are obtained \cite{Prokhorov:2018bql, Buzzegoli:2018wpy}. 

The values calculated in this way did not vanish at the Unruh temperature. We show, that when fourth-order corrections are taken into account in the energy-momentum tensor, this vanishing occurs. Thus, we generalize the result \cite{Becattini:2017ljh} to the case of fermions (at least on the level of energy-momentum tensor of massless fermions), and at the same time confirm that the Unruh effect can be obtained in the Zubarev approach.

We also proposed a formula for energy density in the form of momentum integrals. The motivation for the introduction of this formula is an exact match with the result of the perturbative calculation based on the density operator in the $T>T_U$ region and the correct limit at $a\to 0$. We also derive the energy density using the covariant Wigner function \cite{Becattini:2013fla}. The proposed integral representation can be considered as a modification of this formula resulting from the Wigner function.

\section{Unruh effect from density operator}
\label{Sec:Operator}

The most fundamental object describing a medium in a state of local thermodynamic equilibrium is the density operator, introduced by Zubarev \cite{Zubarev, Weert, Buzzegoli:2017cqy, Becattini:2015nva, Buzzegoli:2018wpy, Becattini:2019dxo}. In the case of zero vorticity $\omega^{\mu}=0$, zero chemical potentials $\mu=\mu_5=0$ and global thermodynamic equilibrium with acceleration \cite{Becattini:2012tc, Becattini:2017ljh, Buzzegoli:2018wpy}, this operator is reduced to the next form \cite{Buzzegoli:2017cqy, Becattini:2017ljh, Buzzegoli:2018wpy}
\begin{eqnarray}
\hat{\rho}=\frac{1}{Z}\exp\Big\{-\beta_{\mu}\hat{P}^{\mu}
-\alpha_{\mu} \hat{K}^{\mu}_x
\Big\} \,,
\label{rho}
\end{eqnarray}
where $\hat{P}^{\mu}$ is a four-momentum operator, $\hat{K}^{\mu}_x$ is a boost operator translated to the vector $x^{\mu}$, and
$\alpha^{\mu} = \frac{a^{\mu}}{T}$ is the vector of thermal acceleration, which is proportional to the usual kinematic acceleration vector in the case of global equilibrium.

In the fourth order of perturbation theory, the operator (\ref{rho}) leads to the energy-momentum tensor of the following form
\begin{eqnarray}
\langle \hat{T}^{\mu\nu}\rangle &=& (\rho_0 +A_1 a^2 T^2  +A_2 a^4 )u^{\mu}u^{\nu}-
(p_0+B_1 a^2 T^2 + B_2 a^4)\Delta^{\mu\nu} \nonumber \\
&&+(A_3 T^2+A_4 a^2)a^{\mu}a^{\nu}+\mathcal{O}(a^6) \qquad
\Delta^{\mu\nu}=g^{\mu\nu}-u^{\mu}u^{\nu}\,,
\label{T}
\end{eqnarray}
where, as before, we assume $a=\sqrt{-a_{\mu}a^{\mu}}$, and $\rho_0, p_0, A_1, A_2, A_3, A_4, B_1, B_2$ are coefficients to be defined. The formula (\ref{T}) is the most general expression allowed by parity requirements in the fourth order of perturbation theory (look \cite{Becattini:2019poj}). In \cite{Buzzegoli:2017cqy} coefficients up to the second order of the perturbation theory were calculated. In the chiral limit $m\to 0$ and at $\mu=\mu_5=0$
\begin{eqnarray}
\rho_0=\frac{7\pi^2 T^4}{60}\,,\, A_1=\frac{1}{24}\,,\,p_0=\frac{\rho_0}{3}=\frac{7\pi^2 T^4}{180}\,,\,B_1=\frac{A_1}{3}=\frac{1}{72}\,,\,A_3=0\,.
\label{coef2}
\end{eqnarray}
Obviously, in the second order of the perturbation theory, the energy-momentum tensor does not vanish. For example, the corresponding expression for energy density (index $Den$ means that the value is calculated using the density operator)
\begin{eqnarray} 
\rho_{Den} = \frac{7 \pi ^2 T^4}{60}+\frac{T^2 a^2}{24}+\mathcal{O}( a^4)
\,,
\label{enden second}
\end{eqnarray}
and it is obvious that it is always positive and does not vanish at $T=T_U$.

We show that taking into account the fourth order corrections leads to that $\langle \hat{T}^{\mu\nu}\rangle(T=T_U)=0$ (third order corrections are forbidden by parity).
Next, we follow the algorithm for calculating corrections in thermal vorticity in quantum statistical averages using the density operator (\ref{rho}), described in  \cite{Buzzegoli:2017cqy, Buzzegoli:2018wpy}, and also in \cite{Prokhorov:2018bql}. So, for  $A_2$ we have
\begin{eqnarray} 
A_2 = \frac{1}{4!}\int_0^{|\beta|} d\tau_xd\tau_yd\tau_zd\tau_f
\langle
T_{\tau} \hat{K}^3_{-i\tau_x u}\hat{K}^3_{-i\tau_y u}
\hat{K}^3_{-i\tau_z u} \hat{K}^3_{-i\tau_f u} \hat{T}^{00}(0)
\rangle_{\beta(x),c}\,,
\label{A2 K}
\end{eqnarray}
where the appearance of four boost operators $\hat{K}$ corresponds to an expansion of up to the fourth order perturbation theory, and the operator $ \hat{T}^{00}$ appears, since we calculate the mean value of the energy density. The index $\beta(x),c$ means that averaging is performed by means of the operator (\ref{rho}) with $\alpha^{\mu}=0$ and only connected correlators are taken, $T_{\tau}$ means ordering by inverse temperature, and $|\beta|=\frac{1}{T}$. Expressing $\hat{K}$ through the energy-momentum tensor, we obtain that the coefficients will be expressed in terms of the quantities (compare with \cite{Buzzegoli:2017cqy, Buzzegoli:2018wpy, Prokhorov:2018bql})
\begin{eqnarray} 
&&C^{\alpha_1\alpha_2|\alpha_3\alpha_4
|\alpha_5\alpha_6|\alpha_7\alpha_8|\alpha_9\alpha_{10}|
ijkl} =\int_0^{|\beta|} d\tau_xd\tau_yd\tau_zd\tau_f d^3x d^3y d^3z d^3f \nonumber \\
&&\times x^{i} y^{j} z^{k} f^{l}\langle
T_{\tau} \hat{T}^{\alpha_1\alpha_2}(\tau_x,\bold{x})\hat{T}^{\alpha_3\alpha_4}(\tau_y,\bold{y})\hat{T}^{\alpha_5\alpha_6}(\tau_z,\bold{z})
\hat{T}^{\alpha_7\alpha_8}(\tau_f,\bold{f})\hat{T}^{\alpha_9\alpha_{10}}(0)
\rangle_{\beta(x),c}\,.
\label{C}
\end{eqnarray}
Then we obtain for the coefficients arising in the fourth order of perturbation theory
\begin{eqnarray} 
A_2 = \frac{1}{4!}C^{00|00|00|00|00|3333}\,,\,B_2 = \frac{1}{4!}C^{00|00|00|00|33|2222}\,,\,A_4 =-B_2+ \frac{1}{4!}C^{00|00|00|00|33|3333}
\,.
\label{coef4}
\end{eqnarray}
The calculation of $C$ differs from the similar calculation in \cite{Buzzegoli:2017cqy, Buzzegoli:2018wpy, Prokhorov:2018bql} only by the larger number of operators under the averaging. We give the final answer right away
\begin{eqnarray} 
A_2 = -\frac{17}{960\pi^2}\,,\quad B_2=\frac{A_2}{3} =-\frac{17}{2880\pi^2}\,,\quad A_4 =0\,.
\label{coef4 result}
\end{eqnarray}
Now it is easy to see that the energy-momentum tensor (\ref{T}) with coefficients (\ref{coef2}) and (\ref{coef4 result}) vanishes at  $T_U$
\begin{eqnarray} 
\langle \hat{T}^{\mu\nu}\rangle=0 \qquad (T=T_U)\,.
\label{main}
\end{eqnarray}
In particular, the energy density takes the form
\begin{eqnarray} 
\rho_{Den} &=& \frac{7 \pi ^2 T^4}{60}+\frac{T^2 a^2}{24} -\frac{17a^4}{960\pi^2}+\mathcal{O}( a^6) \nonumber \\
&=&\frac{1}{240}\Big(T^2-\Big(\frac{ a}{2\pi}\Big)^2\Big)(17 a^2+28\pi^2 T^2)+\mathcal{O}( a^6) 
\,,
\label{enden full}
\end{eqnarray}
and goes to zero at $T=T_U$. The pressure, as it should be, satisfies the equation of state for massless particles $p=\frac{\rho}{3}$ and is also equal to 0 at $T=T_U$.

Other observables are zero due to parity requirements. In particular, in the case of zero vorticity, which we consider, the axial current turns out to be zero - this follows from the direct calculation \cite{Prokhorov:2018bql}, as well as from parity considerations. When calculating the coefficients (\ref{coef2}) and (\ref{coef4 result}) the Belinfante energy-momentum tensor was used, which corresponds to the spin-tensor  $S^{\lambda,\mu\nu}=0$ equal to zero. However, the canonical spin tensor is also zero - this is evident from the fact that it is expressed through an axial current. The vector current is also equal to zero, since we consider the simplest case of $\mu=0$, and the vector current is an odd quantity in chemical potential.

The equality to zero of the observables corresponds to the Minkowski vacuum. Thus, from (\ref{main}) it follows that the Minkowski vacuum corresponds to the proper temperature $T_U=\frac{a}{2\pi}$, which is the essence of the Unruh effect.

\section{Integral representation and Wigner function}
\label{Sec:int}

The formula  (\ref{enden full}) has the form of a polynomial with rather unusual numerical coefficients obtained as a result of calculating quantum correlators of the form (\ref{C}). However, it can be shown that these coefficients can be obtained naturally from integrals with Fermi and Bose distributions. One can check that in the region $T>T_U$ equality
\begin{eqnarray} 
\rho &=& \frac{7 \pi ^2 T^4}{60}+\frac{T^2 a^2}{24} -\frac{17a^4}{960\pi^2} = 2 \int \frac{d^3 p}{(2\pi)^3}\Big(\frac{|\bold{p}|+ia}{1+e^{\frac{|\bold{p}|}{T}+\frac{ia}{2T}}}+\frac{|\bold{p}|-ia}{1+e^{\frac{|\bold{p}|}{T}-\frac{ia}{2T}}}\Big) \nonumber \\
&&+4 \int \frac{d^3 p}{(2\pi)^3}\frac{|\bold{p}|}{e^{\frac{2\pi |\bold{p}|}{a}}-1}
\qquad (T>T_U)\,,
\label{hyp2}
\end{eqnarray}
is exactly satisfied. 

The uniqueness of the integral representation (\ref{hyp2}) shall be subject of special investigation. However, it looks natural and simple. Also this representation leads to the correct result in the limit $a\to 0$. In this case, only the first term in (\ref{hyp2}) remains, which for $a=0$ has the form of the integral of the product of particle energy and the Fermi distribution, as it should be, according to statistical physics. 

On the other hand, (\ref{hyp2}) obtains a rationale from the point of view of another approach, based on the covariant Wigner function for particles with spin $1/2$ \cite{Becattini:2013fla} (see also \cite{Florkowski:2018myy, Florkowski:2018ahw, Prokhorov:2018bql, Florkowski:2018fap, Prokhorov:2017atp, Prokhorov:2018qhq}). In particular, according to (\ref{hyp2}), there is a substitution $\mu\to\mu\pm \frac{ia}{2}$, which means that the acceleration plays the role of an imaginary chemical potential, which was previously also observed for axial current in the approach with the Wigner function \cite{Prokhorov:2018qhq}.

Using the Wigner function from \cite{Becattini:2013fla}, one can calculate the mean values of different observables. In particular, in \cite{Prokhorov:2017atp} and \cite{Prokhorov:2018qhq} a method for obtaining exact non-perturbative results is described using the example of axial current. In our case, it is necessary to calculate the energy density. According to \cite{Prokhorov:2017atp}, it is necessary to decompose the energy density twice into a Taylor series
\begin{eqnarray} \nonumber
\rho_{Wig}&=&\frac{1}{2}\int \frac{d^3 p}{(2\pi)^3}\varepsilon \Big(\sum^{\infty}_{n=0}(-1)^n \exp\big[t(n+l)(\beta\cdot p -\xi)\big] \\
&&\times\sum^{\infty}_{m=0}\frac{1}{m!} \big(-\frac{1}{2}t(n+l)\big)^{m}\mathrm{tr}\Big[\big(\varpi :\Sigma\big)^{m}\Big]+\big(\xi\to-\xi,\varpi\to-\varpi \big)\Big)\,,
\label{Wigner calc Taylor}
\end{eqnarray}
where index $Wig$ means that the value is calculated using the Wigner function \cite{Becattini:2013fla}, $\varpi_{\mu\nu}= -\frac{1}{2}(\partial_{\mu}\beta_{\nu}-\partial_{\nu}\beta_{\mu})$ is thermal vorticity tensor, $\xi=\frac{\mu}{T}$ is chemical potential divided by temperature,  $t=1$, $l=0$ or $t=-1$, $l=1$, and in brackets there is the contribution of antiparticles, distinguished by signs of chemical potential and vorticity. Next, one needs to calculate a trace in each term of the series and sum up them back. As a result, we obtain the following expression for the energy density
\begin{eqnarray}
\rho_{Wig} =2 \int \frac{d^3 p}{(2\pi)^3}\varepsilon \Big(\frac{1}{1+e^{\frac{\varepsilon}{T}+\frac{i a}{2T}}}+\frac{1}{1+e^{\frac{\varepsilon}{T}-\frac{i a}{2T}}}\Big)\,,
\label{W finite m}
\end{eqnarray}
where the result is given for the case of $\xi=0$ and global thermodynamic equilibrium. As far as we know, the result (\ref{W finite m}) is new and has never appeared before.

The form of the expression (\ref{W finite m}) is the motivation for the integral representation (\ref{hyp2}). Note that in  (\ref{W finite m}) the condition $\rho(T=T_U)=0$, cannot be achieved, unlike (\ref{hyp2}). This fact was previously shown in \cite{Florkowski:2018myy}, where the Boltzmann limit was investigated  \footnote{We are grateful to W. Florkowski and E. Speranza for the discussion.}. Apparently, this indicates the limited possibility of using the Wigner function \cite{Becattini:2013fla}  in describing the effects associated with acceleration (while it works well for the effects of vorticity, as was shown in \cite{Prokhorov:2018bql}). In (\ref{hyp2}) the condition $\rho(T=T_U)=0$ is ensured by adding the second integral with the Bose distribution with a temperature equal to the Unruh temperature. In general, the integral representation (\ref{hyp2}) can be considered as a modification of the formula (\ref{W finite m}) obtained from the Wigner function.

Let's discuss the properties of energy density, following from (\ref{hyp2}). Integrals in (\ref{hyp2}) lead to an additional contribution at $T<T_U$, containing the function $\lfloor \cdot \rfloor$, which takes the integer part
\begin{eqnarray} 
\rho &=&\frac{7 \pi ^2 T^4}{60}+\frac{T^2 a^2}{24} -\frac{17a^4}{960\pi^2}+\Big\lfloor\frac{1}{2}+\frac{a}{4\pi T} \Big\rfloor
\Big(\frac{\pi T^3 a}{3}+\frac{T a^3}{4 \pi}\Big) \nonumber \\
&&- \Big\lfloor\frac{1}{2}+\frac{a}{4\pi T} \Big\rfloor^2
\Big(\frac{T^2 a^2}{2}+2\pi^2 T^4\Big)-
\frac{4 \pi T^3 a}{3}\Big\lfloor\frac{1}{2}+\frac{a}{4\pi T} \Big\rfloor^3
+4 \pi^2 T^4 \Big\lfloor\frac{1}{2}+\frac{a}{4\pi T} \Big\rfloor^4
\,.
\label{hyp2 integrated}
\end{eqnarray}
The appearance of an additional contribution at $T<\frac{a}{2 \pi}$, does not contradict (\ref{enden full}), since the formula  (\ref{enden full}) was obtained in the framework of the perturbative approach. Note that the contributions with the integer part at $T<\frac{a}{2 \pi}$ also appeared in the axial current in the approach with the Wigner function \cite{Prokhorov:2018qhq}.

The formula (\ref{hyp2}) admits an interesting interpretation. In the first term, describing fermions, the effects of motion of the medium lead to the appearance of an imaginary contribution to the energy $\varepsilon\to\varepsilon\pm ia$ and the chemical potential  $\mu\to\mu\pm i\frac{a}{2}$. The latter leads to the fact that at $T=T_U/(2k+1)\,,\,k=0,1..$ in (\ref{hyp2 integrated}) instabilities arise, namely, discontinuities appear in second, third and fourth order derivatives with respect to temperature $\frac{\partial^2\rho}{\partial T^2}$, $\frac{\partial^3\rho}{\partial T^3}$,  $\frac{\partial^4\rho}{\partial T^4}$ (similarly to axial current in \cite{Prokhorov:2018qhq}). The instability at $T=T_U$, can be considered as a manifestation of the Unruh effect.

\begin{figure*}[!h]
\centerline{\includegraphics[width=8.6 cm]{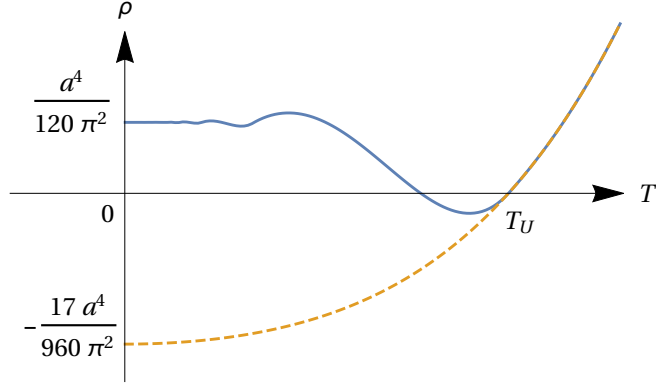}}\vspace{1cm}
\caption{Energy density as a function of temperature. The proposed formulas  (\ref{hyp2}) and (\ref{hyp2 integrated}) correspond to the solid blue line; the perturbative result (\ref{enden full}), obtained from the density operator, is shown by the dashed orange line.}
\label{fig:plot}
\end{figure*}

The appearance of an additional contribution with the integer part in  (\ref{hyp2 integrated}) leads to the fact that below $T_U$ the behavior of (\ref{enden full}) and (\ref{hyp2 integrated}) is different, which is shown in Fig.\ref{fig:plot}. Formula (\ref{hyp2 integrated}) in contrast to (\ref{enden full}) leads to a non-monotonic oscillating behavior, which is associated with additional terms with an integer part. However, according to \cite{Becattini:2017ljh} the Unruh temperature has to be considered as minimal, and the region $T<T_U$ is forbidden.

Note, in conclusion, that according to (\ref{hyp2 integrated}), when $T>T_U$ the maximum power of acceleration in the energy density is four. This is also supported by the fact that, starting with the sixth order in acceleration, negative powers of temperature would have arisen, which is necessary to preserve the correct dimension. Thus, we would expect that the perturbative result (\ref{enden full}) is exact at $T>T_U$ and all the corrections above $T_U$ are zero, although this fact requires more rigorous justification.

\section{Conclusions}
\label{Sec:conclusion}

We have shown on the basis of Zubarev density operator that in the fourth order of perturbation theory, the energy-momentum tensor of inertial observer vanishes at the Unruh temperature of comoving observer. Also in the case under consideration of zero chemical potentials and zero vorticity, the spin tensor and the vector and axial currents are also equal to zero. Thus, the Minkowski vacuum is visible to the accelerated observer as a medium filled with a thermal bath with a temperature $T_U=\frac{a}{2\pi}$, which is the essence of the Unruh effect and generalizes the results of \cite{Becattini:2017ljh} to the case of fermions.

We present the obtained perturbative result in the form of momentum integrals. The proposed formula exactly coincides with the perturbative result at  $T>T_U$, and can be motivated by a formula derived from the Wigner function.

{\bf Acknowledgements}
The reported study was funded by RFBR according to the research project 18-02-40056.


\end{document}